\documentclass[aps,prx,superscriptaddress,reprint,amsmath]{revtex4-1}

\usepackage{graphicx, color}
\usepackage{amsfonts,amssymb,amsmath}

\begin{document}

\title{Gigantic negative magnetoresistance in a disordered \\
topological insulator}

\author{Oliver Breunig}
\thanks{These authors contributed equally to this work}
\author{Zhiwei Wang}
\thanks{These authors contributed equally to this work}
\affiliation{Physics Institute II, University of Cologne, Z\"ulpicher Str. 77, 50937 K\"oln, Germany}
\author{A. A. Taskin}
\affiliation{Physics Institute II, University of Cologne, Z\"ulpicher Str. 77, 50937 K\"oln, Germany}
\author{Jonathan Lux}
\author{Achim Rosch}
\affiliation{Institute for Theoretical Physics, University of Cologne, Z\"ulpicher Str. 77, 50937 K\"oln, Germany}
\author{Yoichi Ando}
\affiliation{Physics Institute II, University of Cologne, Z\"ulpicher Str. 77, 50937 K\"oln, Germany}

\begin{abstract}
With the recent discovery of Weyl semimetals, the phenomenon of negative magnetoresistance (MR) is attracting renewed interest. While small negative MR can occur due to the suppression of spin scattering or weak localization, large negative MR is rare in materials, and when it happens, it is usually related to magnetism. The large negative MR in Weyl semimetals is peculiar in that it is unrelated to magnetism and comes from chiral anomaly. Here we report that there is a new mechanism for large negative MR which is not related to magnetism but is related to disorder. In the newly-synthesized bulk-insulating topological insulator TlBi$_{0.15}$Sb$_{0.85}$Te$_2$, we observed gigantic negative MR reaching 98\% in 14 T at 10 K, which is unprecedented in a nonmagnetic system. 
Supported by numerical simulations, we argue that this phenomenon is likely due to the Zeeman effect on a barely percolating current path formed in the disordered bulk. Since disorder can also lead to non-saturating linear MR in Ag$_{2+\delta}$Se, the present finding suggests that disorder engineering in narrow-gap systems is useful for realizing gigantic MR in both positive and negative directions.
\end{abstract}
\maketitle

Magnetic fields tend to localize electrons by trapping them in cyclotron orbits, leading to positive magnetoresistance (MR) in metals and semiconductors. Large {\it negative} MR is therefore unusual and is always a signature of some peculiar physics. So far, three mechanisms are widely known to cause large negative MR: (i) quenching of the Kondo effect \cite{Andrei1983,Hanaki2001}, (ii) magnetic-field-induced phase transition from a paramagnetic insulator to a ferromagnetic metal (so-called colossal MR) \cite{RamirezCMR}, and (iii) chiral anomaly in Weyl semimetals \cite{NIELSEN1983, Aji2012, Son2013, Burkov2014}. The last one is expected to be observed only in the longitudinal configuration (i.e. when the current is parallel to the magnetic field) and has been of significant interest since three-dimensional (3D) Dirac semimetals and Weyl semimetals started to attract attentions \cite{Wan2011, Young2012}. Intriguingly, large negative, longitudinal MR of similar character has been observed in materials having no Weyl nodes \cite{Kikugawa2016}, and the meaning of negative MR in nonmagnetic systems is currently being scrutinized \cite{Goswami2015,Arnold2016}. In this context, finding of a new mechanism for large negative MR in zero-gap or narrow-gap systems without magnetism would be important for establishing general understanding of the negative MR. This paper reports an unexpected discovery of gigantic negative MR in a nearly-bulk-insulating topological insulator (TI), which turned out to be a platform to realize large negative MR in the bulk through a new mechanism related to disorder.

The key feature of 3D TIs is the topologically-protected surface states \cite{Hasan2010,Ando2013}, and significant efforts have been devoted to find ways to achieve sufficiently bulk-insulating samples to address the transport properties through the surface states \cite{Ando2013,Ren2010}. The most successful strategy to achieve this goal has been to adopt the concept of compensation, namely, balancing the numbers of donors and acceptors \cite{Ren2011}. This has led to the successful achievements of surface-dominated transport in bulk single crystals of Bi$_{2-x}$Sb$_x$Te$_{3-y}$Se$_y$ (BSTS) \cite{Taskin2011}. We have recently synthesized another solid-solution system of 3D-TIs, TlBi$_x$Sb$_{1-x}$Te$_2$ (TBST), which belongs to the family of thallium-based ternary III-V-VI$_2$ chalcogenides \cite{Lin2010,Yan2010,Eremeev2011,Sato2010,Chen2010} and the chemical potential can be tuned through the bulk band gap between $n$- and $p$-type states \cite{Trang2016}. We found that the maximal compensation in TBST is achieved near $x \simeq$ 0.2 (see Supplementary Information for details), where one can obtain nearly bulk-insulating samples. The gigantic MR is discovered in such a compensated sample of TBST.

\section{Results}

\begin{figure}[b]
\centering
\includegraphics[width=\columnwidth]{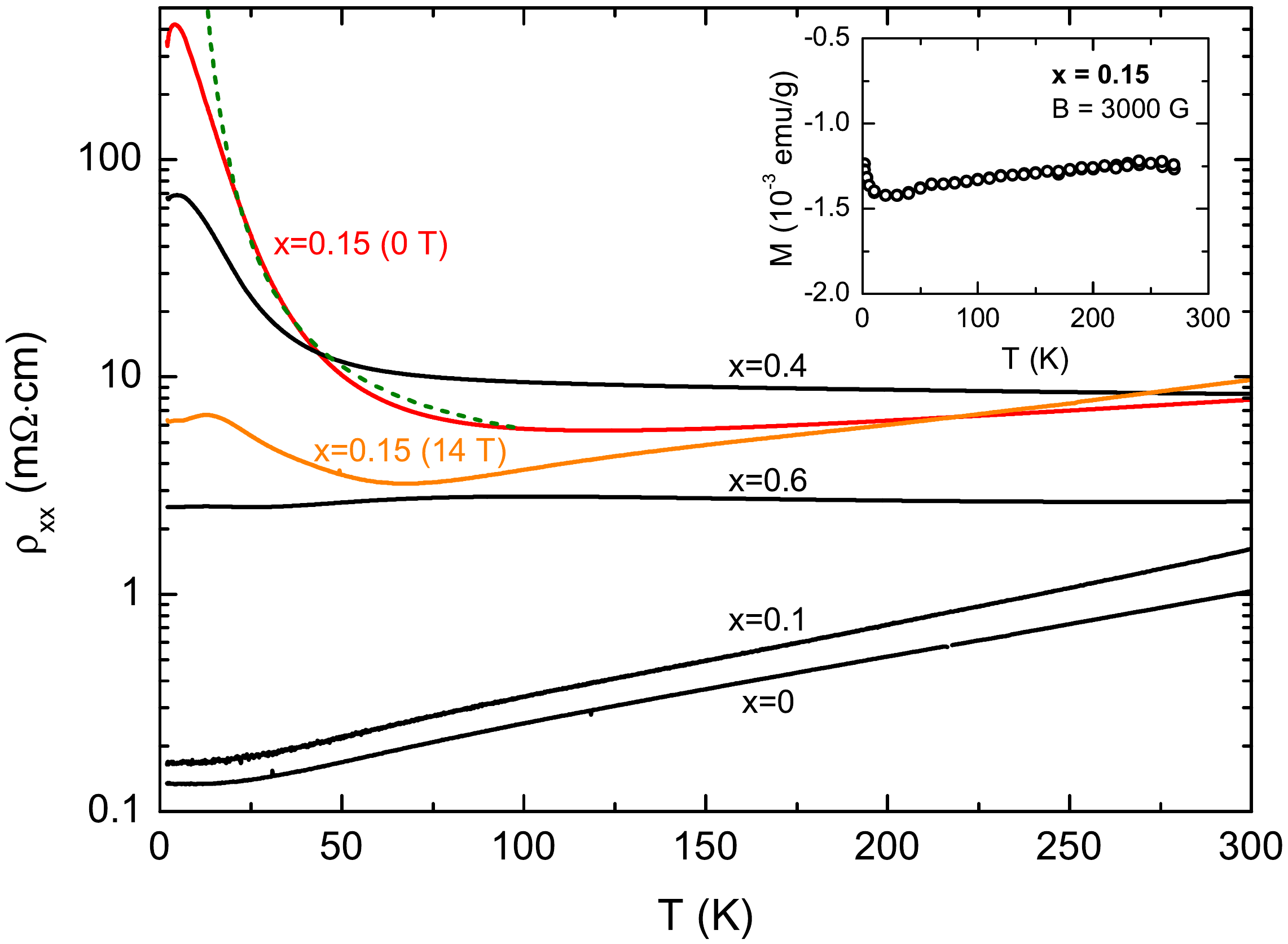}
\caption{{\bf Resistivity behavior of TlBi$_x$Sb$_{1-x}$Te$_2$.} The temperature dependencies of $\rho_{xx}$ in the $ab$ plane are plotted for $x$ = 0, 0.1, 0.15, 0.4, and 0.6 in 0 T; the green dashed line depicts a simple activated behavior with an activation energy of 6 meV. The data in 14 T are additionally shown for $x$ = 0.15 (orange line).  Inset shows the magnetization data for $x$ = 0.15.
} 
\label{fig:1}
\end{figure}

\begin{figure}[b]
\centering 
\includegraphics[width=\columnwidth]{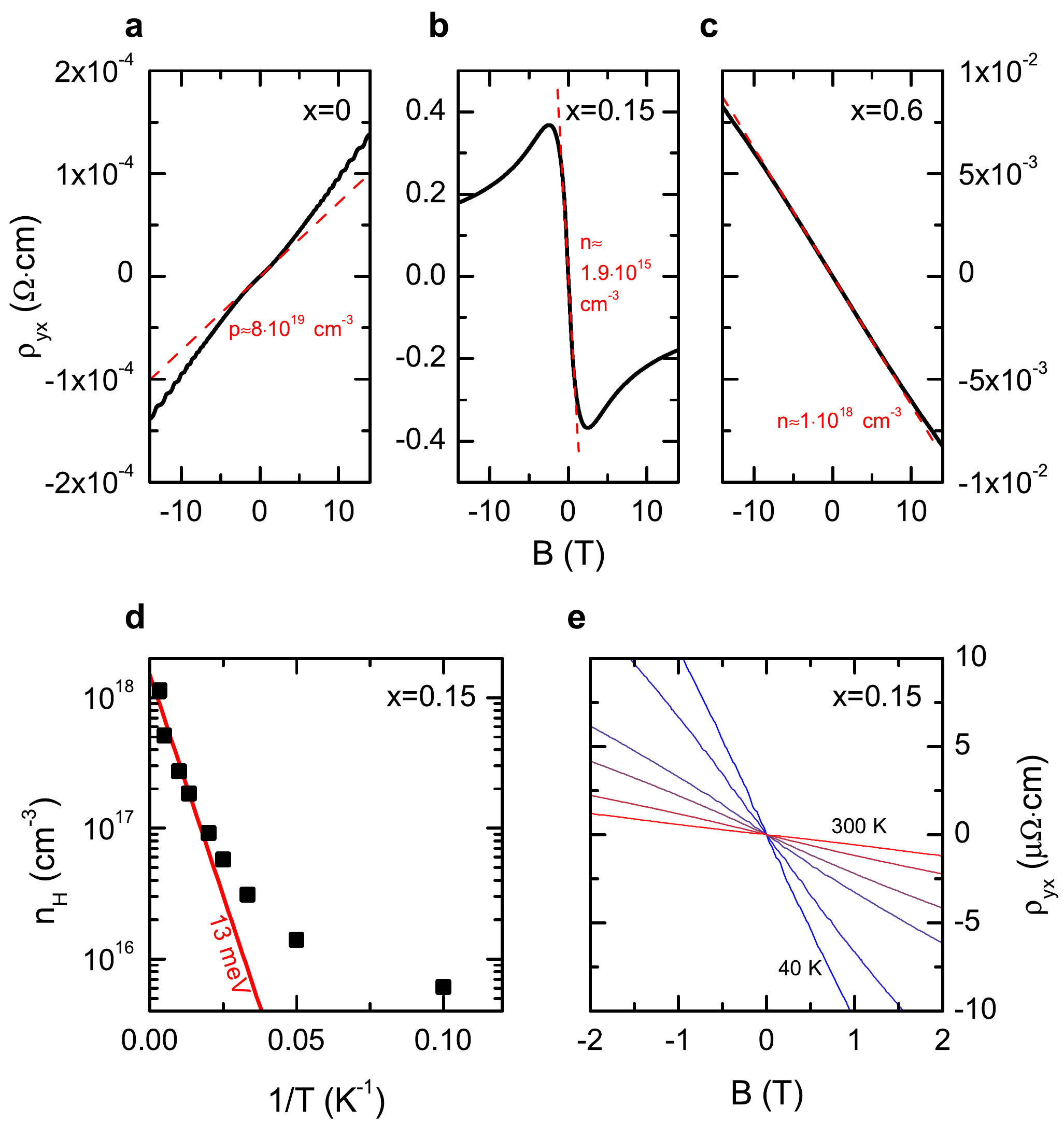}
\caption{{\bf Hall effect.} (a-c) $\rho_{yx}(B)$ data for $x$ = 0, 0.15, and 0.6 obtained at 2 K; red dashed lines represent the low-field slope yielding the indicated densities of holes ($p$) and electrons ($n$). The strongly changing slope in panel (b) might prompt a two-band analysis, but the gigantic negative MR clearly negates its applicability; according to our puddle scenario, the non-linear $\rho_{yx}(B)$ comes form a change in the number of mobile bulk carriers with $B$.
(d) Arrhenius plot of the Hall carrier density $n_{\rm H}$ for $x$ = 0.15, giving an approximate activation energy of 13 meV (red line).
(e) $\rho_{yx}(B)$ data for $x$ = 0.15 measured at $T$ = 300, 200, 100, 75, 50, and 40 K. } 
\label{fig:2}
\end{figure}

Figure ~\ref{fig:1} shows the temperature dependencies of $\rho_{xx}$ in TBST for various $x$ values. Among the samples showing metallic behavior, $x$ = 0 and 0.1 are $p$-type, while $x$ = 0.6 is $n$-type [see Figs. \ref{fig:2}a and \ref{fig:2}c for the Hall data]. A slight increase in $x$ from 0.1 to 0.15 causes $\rho_{xx}$ at $\sim$10 K to increase by three orders of magnitude to 0.4 $\Omega$cm, and an insulating behavior ($\partial \rho_{xx}/\partial T<0$) arises between 10 to 100 K. We focus on this $x=0.15$ sample, in which the Hall carrier density $n_\mathrm{H}$ extracted from the zero-field slope of $\rho_{yx}(B)$ (Fig.~\ref{fig:2}b) is only 1.9 $\times$ 10$^{15}$ cm$^{-3}$ at 2 K and approximately follows the Arrhenius law with an activation energy $\Delta_\mu$ of 13 meV [Fig.~\ref{fig:2}d]. The compensation in TBST is not as effective as that in BSTS \cite{Taskin2011} and the maximum attainable resistivity is lower; consequently, we found no evidence for a surface contribution to the transport properties.

\begin{figure}[b]
\centering
\includegraphics[width=\columnwidth]{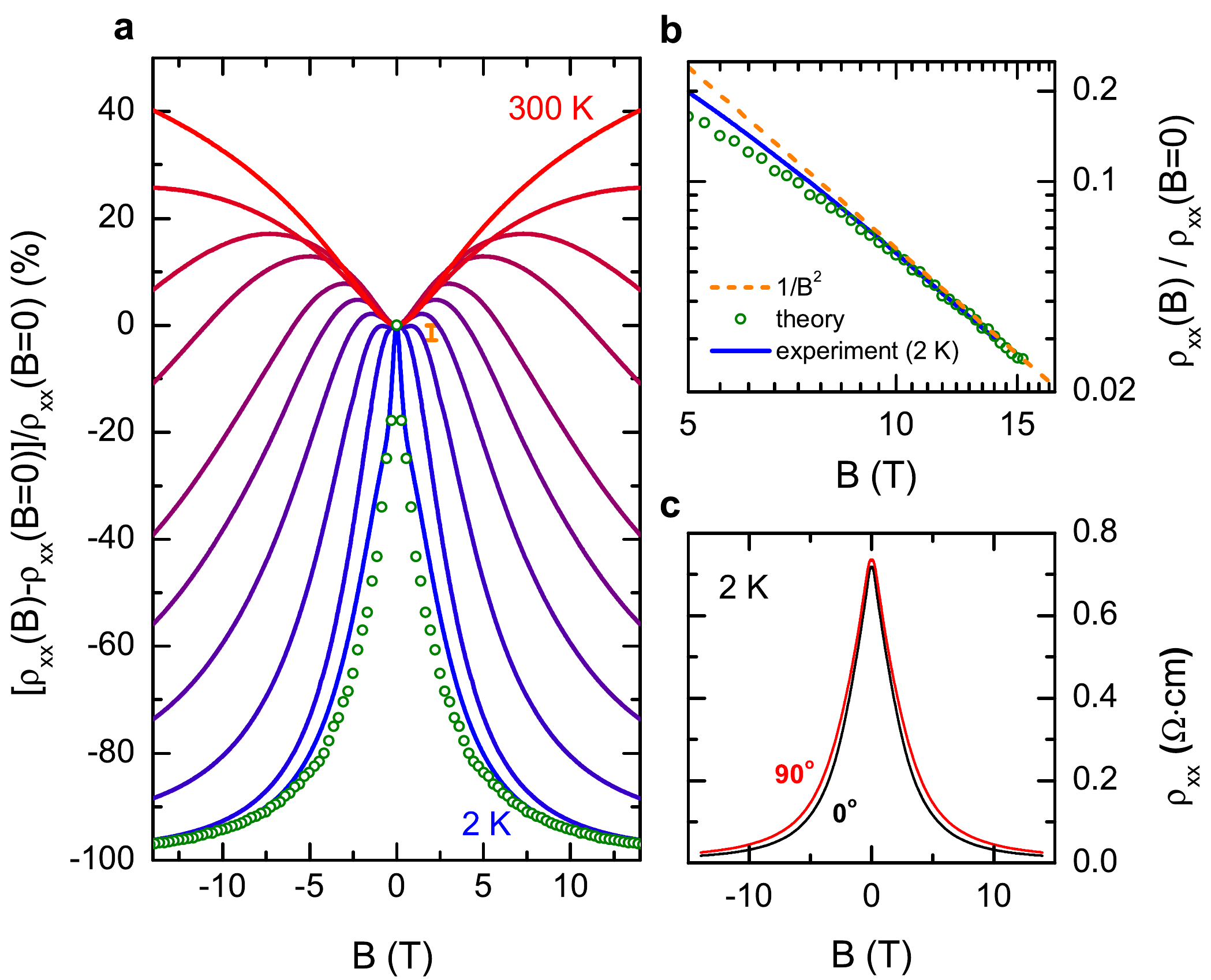}
\caption{{\bf Magnetoresistance of TlBi$_{0.15}$Sb$_{0.85}$Te$_2$ in magnetic fields along the $c$ axis.} 
(a) Relative change of $\rho_{xx}(B)$ with respect to $\rho_{xx}(B=0)$ at different constant temperatures $T$ = 300, 200, 100, 75, 50, 40, 30, 20, 10, and 2 K from top (red) to bottom (blue); open green circles are obtained by numerically solving a resistor network for the disordered Coulomb system at $T$ = 0 K as explained in the main text. The vertical bar at 2 T shows an estimate of the possible 3D weak localization contribution \cite{Altshuler}.
(b) Double-logarithmic plot of $\rho_{xx}(B)/\rho_{xx}(B=0)$ showing that the low-$T$/ high-$B$ data approach a $1/B^2$ dependence indicated by a dashed orange line; the numerical results (open green circles) also approach the $1/B^2$ behavior at high $B$. 
(c) $\rho_{xx}(B)$ data for two different magnetic field orientations, $B\parallel c$ and $j\perp B\perp c$; here, the $c$-axis is along the [111] direction of the rhombohedral unit cell and is normal to the crystallographic layers. } 
\label{fig:3}
\end{figure}

The surprising finding is that the magnetic field of 14 T leads to a reduction of $\rho_{xx}$ at 10 K by nearly two decades from 400 to 6.5 m$\Omega$cm (orange line in Fig.~\ref{fig:1}). How this gigantic negative MR evolves with temperature is shown in Fig.~\ref{fig:3}a; with decreasing temperature, the MR changes from entirely positive at $\gtrsim 200$ K to entirely negative at $\lesssim 20$ K. The additional cusp-like feature seen at 2 K for $B \lesssim$ 1 T may be attributed to a weak localization effect, but it cannot explain the large negative MR; this can be seen by the estimate of the possible 3D weak localization contribution \cite{Altshuler} for 2 T shown in Fig. 3a (see Supplementary Information for details). The logarithmic plot of $\rho_{xx}(B)$ shown in Fig.~\ref{fig:3}b suggests that the MR behavior approaches a $1/B^2$ dependence in high fields at low temperature.

It is important to note that the compositional analysis using inductively-coupled plasma atomic-emission spectroscopy found no magnetic impurities in TBST, which is corroborated by the magnetic susceptibility data (inset of Fig.~\ref{fig:1}). Hence, magnetism is not involved in this gigantic negative MR. Also, the chiral anomaly is irrelevant, because the negative MR shows up in the transverse configuration. Its origin is probably not related to the anisotropy in the band structure, because the difference between $B\parallel c$ and $B\perp c$ is negligible (Fig.~\ref{fig:3}c).

Phenomenologically, the appearance of the negative MR in TBST is tied to the emergence of an insulating behavior. In fact, among the investigated samples, all those that show an insulating behavior ($0.13 \le x \le 0.4$) presented a negative MR which was most pronounced in the $x$ = 0.15 sample shown here (additional data for other $x$ values are shown in the Supplementary Information). The onset temperature of the negative MR upon cooling compares well with the onset of an activated behavior in $\rho_{xx}(T)$. A fit of the data below 100 K to $\rho_{xx} \propto e^{E_A/k_\mathrm{B} T}$, shown as a dashed green line in Fig.~\ref{fig:1}, yields an activation energy $E_A \approx$ 6 meV. This value is substantially smaller than the bulk band gap of $\sim$0.1 eV \cite{Chen2010,Trang2016}. Such a small effective activation energy in compensated TIs \cite{Ren2010,Ren2011} has been discussed to be a signature of the formation of charge puddles \cite{Chen2013, Skinner2013, Skinner2012}, which we discuss in the next section to be responsible for the gigantic negative MR.

\section{Discussions}

{\bf Electron puddles. }
The concept of electron and hole puddle formation has been adopted in the framework of compensated TIs to explain their small effective activation energy \cite{Skinner2012,Skinner2013,Chen2013} as well as the unusual temperature dependence of the bulk optical conductivity \cite{Borgwardt2015}. Basically, the strong fluctuations of the Coulomb potential introduced by positively charged empty donors and negatively charged occupied acceptors, which coexist in compensated systems, inevitably create regions where the conduction-band bottom or the valence-band top cross the chemical potential, leading to the appearance of locally conducting puddles containing either electrons or holes \cite{Shklovskii1972}. 

Similar to other compensated TIs, the impurity states in TBST are shallow due to a rather large dielectric constant $\epsilon \approx 200$, approximated from that of TlSbTe$_2$ \cite{Deger2015}. Thus, their energies are close to the edge of the conduction or valence band. The $x$ = 0.15 sample is $n$-type and is slightly away from complete compensation, leading to the pinning of the chemical potential at low to moderate temperature to the donor impurity levels, which lie below but close to the conduction band bottom. The obtained activation energy $\Delta_{\mu}$ = 13 meV corresponds directly to $E_0 - \mu$, which is the energy difference between the conduction band bottom in 0 T, $E_0$, and the chemical potential $\mu$. The dominance of a single $n$-type channel at moderate temperature is also indicated by the linear $\rho_{yx}(B)$ behavior at $T \ge$ 40 K (Fig.~\ref{fig:2}e).  

At low temperature where the screening due to thermally activated carriers is negligible, the fluctuations of the Coulomb potential arising from charged impurities pile up. In this situation, the same mechanism as in the case of electron-hole puddles \cite{Skinner2012} enforces the formation of spatially separated electron puddles. Skinner, Chen, and Shklovskii \cite{Skinner2012} pointed out that one obtains effectively a reduced activation energy in the presence of strong fluctuations of the Coulomb potentials. Hence, the activation energy $E_A$ ($\approx$ 6 meV) observed in the low-temperature $\rho_{xx}(T)$ data is actually expected to differ from $\Delta_\mu$ extracted from the $n_{\rm H}(T)$ behavior at higher temperature. 

Below 4 K, the $\rho_{xx}(T)$ behavior for $x$ = 0.15 is metallic; this is different from the prediction for a compensated TI in Ref. \cite{Skinner2012}, where a variable range hopping transport between puddles was postulated. Furthermore, in the case of hopping conduction, a positive MR would be expected due to the decreasing size of the atomic orbitals in strong magnetic fields \cite{Shklovskii1984, ESbook}. The metallic behavior indicates that either the electron puddles in the slightly doped samples weakly percolate or another conduction mechanism (e.g., via impurity bands) connects them.

\begin{figure*}[t]
\centering
\includegraphics[width=15cm]{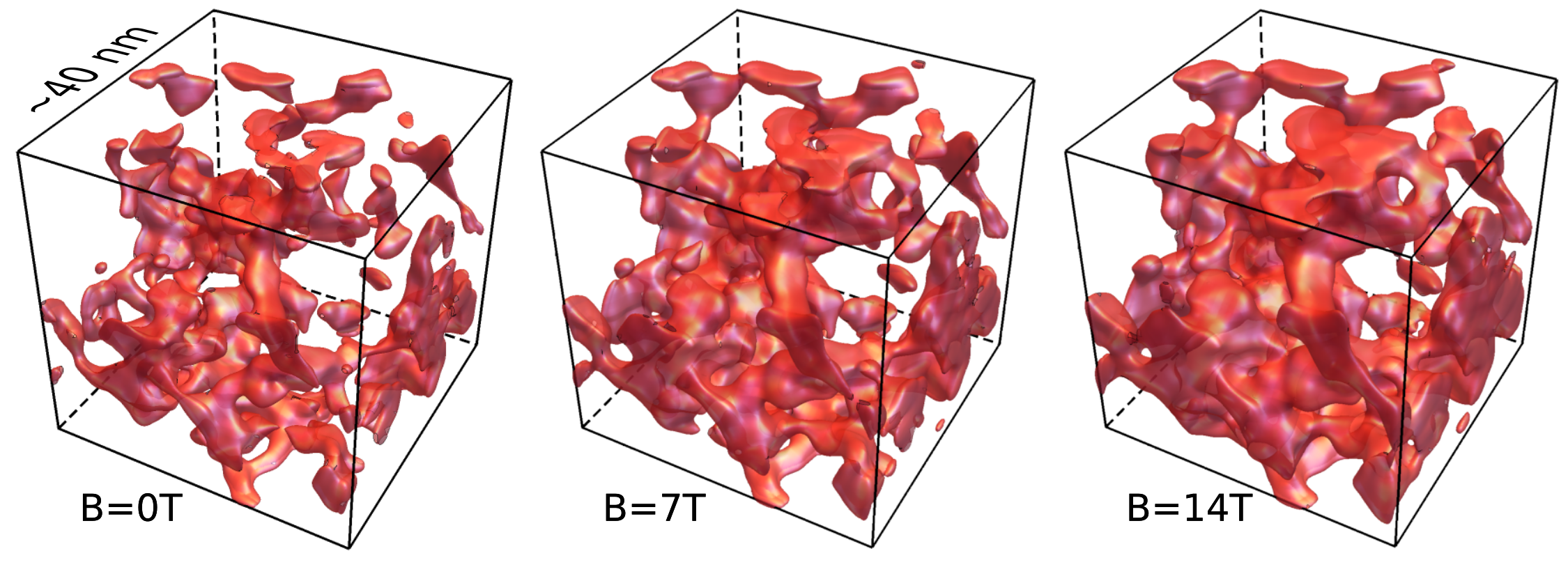}
\caption{{\bf Simulation of electron puddles in magnetic fields.} Spatial distribution of electron puddles in an imperfectly compensated Coulomb system at $T$ = 0 K, simulated for $g$ = 6, $E_c$ = 4.9 meV, and $K$ = 0.95.
The colored surfaces represent where the spatially-fluctuating Coulomb potential $\phi(\bold{r})$ becomes equal to $(E_0 - \mu) - E_Z(B)$, which is the criterion for puddle formation. 
With increasing magnetic field $B \approx$ 0, 7, and 14 T (from left to right), the volume of the enclosed regions increases due to the effect of Zeeman energy. 
The chemical potential is assumed to be constant due to the pinning by the impurity levels, which work as a reservoir of delocalized carriers. 
Simulations are shown for a cube with a width of 40 nm corresponding  to $\sim 3 \times 10^4$ dopants. We have checked that the qualitative behavior does not change in larger systems. 
} 
\label{fig:4}
\end{figure*}

{\bf Numerical Simulations. }
Based on the above picture of the spatially disordered electronic states caused by electron puddles, we propose that the increasing conductivity in applied magnetic fields is a result of an increase of percolating puddles due to the Zeeman energy. To understand this effect, we performed numerical simulations of an imperfectly compensated Coulomb system. 
The characteristic energy scale of such a system is given by the Coulomb energy between neighboring dopants $E_c = e^2/(4 \pi \epsilon \epsilon_0 N^{-1/3})$ \cite{Borgwardt2015}, where $N$ denotes the dopant density. Since the dielectric constant in TBST is large ($\epsilon \approx 200$ \cite{Deger2015}) as discussed above, the temperature scale $E_c/k_B$ becomes small and we use this parameter to adjust the numerical results to the experimental data. 
Also, as the negative MR is most pronounced at low temperature, we restrict the analysis to the case $k_B T \ll E_c$.
For the simulation, the degree of compensation was set at $K=N_A/N_D = 0.9-0.95$, where $N_A$ and $N_D$ denote the densities of acceptors and donors, respectively. 
This implies that the chemical potential is close to the conduction band edge and that puddles are almost exclusively formed by electrons. 
The electron puddles are spatially confined to regions where the position-dependent electrostatic potential $\phi({\bold r})$ (which is responsible for the local band bending) exceeds $E_0 - \mu$; 
in other words, electron puddles appear when the local band bending $\phi({\bold r})$ brings down the conduction band edge below the chemical potential.

Assuming a single spherical band with effective mass $m^*$, the typical density of charges delocalized within the puddles at low temperature is given by
\begin{align}
 n  &= 2 \int \frac{d^3 \boldmath{k}}{(2\pi)^3} \Theta \left( E_c - \frac{\hbar^2 k^2}{2 m^*} \right) \nonumber\\ 
 &= \frac{1}{3 \pi^2 \hbar^3} \left( \frac{2 m^* e^2}{4 \pi \epsilon \epsilon_0}\right)^{3/2} N^{1/2} \,.
\end{align}
For $\epsilon = 200, m^* =0.2 m_e$ and $N = 3 \times 10^{20}\,\mathrm{cm}^{-3}$, we find a density $n \approx 10^{17}\,\mathrm{cm}^{-3}$, which is significantly smaller than that of the dopants, i.~e.~$n \ll N$. 
Thus, the impurity states serve as a reservoir for the delocalized states and approximately fix the chemical potential. 

In such a situation, the volume accessible by the delocalized states (i.e. the volume occupied by the puddles) increases with increasing magnetic field $B$ due to the Zeeman band shift. 
Adopting $g \approx 6$ from a similar system TlBiSSe \cite{Novak2015}, the down shift of the band bottom, $g\mu_\mathrm{B}B$, is estimated to be $0.35 \,\mathrm{meV}/\mathrm{T}$. 
This effect is visualized in Fig.~\ref{fig:4}, which shows a typical simulation result for $K=0.95$ (details are explained in the Methods section). Here we plot the surface contours where the local electrostatic potential supports puddle formation, i.e. where $\phi(\bold{r}) = (E_0 - \mu) - E_{\rm Z}(B)$ is satisfied [$E_{\rm Z}(B) = g \mu_{\rm B}B$ is the Zeeman energy], 
for $E_{\rm Z}/E_c$ = 0, 0.5, and 1, corresponding to $B \simeq$ 0, 7, and 14 T, respectively.
Note that, due to the pinning of the chemical potential to the impurity levels, $E_0 - \mu$ is expected to be unchanged with $B$. Within the enclosed regions (i.e. electron puddles), delocalized states exist. One can clearly see that with increasing magnetic field, previously disconnected puddles merge. Note that the occupation of most impurity states, and therefore the profile of the potential fluctuations, is assumed not to change with magnetic field. 
The increase in the percolating paths in $B >0$ demonstrated in Fig.~\ref{fig:4} naturally leads to an increase in the electrical conductivity. 

To quantify this effect we have calculated the conductivity at different magnetic fields using a random resistor network \cite{Derrida1984} (details are explained in the Methods section).
The local conductivity of each resistor in this network was calculated according to $\sigma (\mathbf{r}) = \mu_e n(\mathbf{r}) $, where we have assumed a constant electron mobility $\mu_e$. For the calculation of the density $n(\mathbf{r})$ we have assumed 
a locally free electron gas which gives $n(\mathbf{r}) \sim \left( \phi(\bold{r}) - (E_0 - \mu) + E_{\rm Z}(B) \right)^{3/2} $ and $\sigma(\mathbf{r})=0$ if $\phi(\bold{r}) < (E_0 - \mu) - E_{\rm Z}(B)$.
Note that this approach is tailored to describe the regime where puddles overlap and start to percolate. In this regime, variable-range hopping effects \cite{Skinner2012} (not included in our resistor network) can be neglected. 
Furthermore, weak localization effects, which we have already shown to be small, are not included.
By fitting the numerical results at $B=0$ to the experimental data, we find a mobility of $\mu_e \approx 11,500\,\mathrm{cm}^2$/Vs corresponding to a scattering rate of $1/\tau \approx 10^{12}/$s, which is in reasonable agreement with the scattering rates measured in similar systems, see Ref. \cite{Borgwardt2015} for an overview. These numbers depend, however, sensitively on $K$ and therefore are only an order of magnitude estimate.
Furthermore, one should note that this number only characterizes the typical mobility {\it within} a puddle, not the effective mobility obtained by averaging over the bulk. The shape and width of $\rho_{xx}(B)/\rho_{xx}(B=0)$ depend only weakly on $K$, while $E_c$ determines the width of the curve and is used as an adjusting parameter; we find $E_c = 60\, \pm 10$ K reproduces the data well for $g=6$. We show in Figs.~\ref{fig:3}a and 3b our numerical results for $\rho_{xx}(B)/\rho_{xx}(B=0)$ with $K=0.9$ and $E_c=56\,$K, which correspond to a dopant densisty $N = 3 \times 10^{20}\,\mathrm{cm}^{-3}$ and is in reasonable agreement with studies of similar systems \cite{Borgwardt2015}. Thus, a set of reasonable parameters reproduces the observed negative MR in our random resistor network simulation. 
Remarkably, our numerics even reproduces the approximate $1/B^2$ asymptotics of $\rho_{xx}$ for high fields, see Fig. 3b. In even higher fields, when either a large volume fraction of the sample becomes a good metal or when our approximation of a fixed chemical potential breaks down, we expect a modified high-field asymptotics.

{\bf Implications. }
The above mechanism for the MR is expected to be independent of whether the configuration is transverse or longitudinal. Indeed, as is shown in the Supplementary Information, essentially the same negative MR is observed in the longitudinal configuration. As for the temperature dependence, although our numerical calculations were performed only for 0 K, it is naturally expected that the random Coulomb potential is gradually screened with increasing temperature by thermally-activated carriers, which results in smearing of the puddles; consequently, the negative MR is gradually diminished. The positive MR observed at high temperature is relatively large and would be an interesting topic for future studies.
 
It is prudent to mention that large negative MR has not been observed in other compensated TI materials such as BSTS. Our numerical simulations show that for the occurrence of the large negative MR, it is important that the parameter $K$ is at a right value away from 1.0, which means that the compensation is not perfect and the puddles of only one carrier type is weakly percolating in 0 T. Obviously, such a situation is not realized in BSTS, in which the degree of compensation is better than that in the present TBST series. We also note that the density of naturally created vacancies in the crystals may widely vary among different compounds, resulting in significantly different landscape/density of puddles and hence different percolation behavior. 

As our simulations suggest, the gigantic negative MR in TlBi$_{0.15}$Sb$_{0.85}$Te$_2$ is likely due to a new mechanism related to the spatial disorder of the electronic system in the form of puddles, which results in magnetic-field-sensitive percolation of the current paths. In this regard, it is interesting that large positive, non-saturating linear MR has been shown to originate from spatially distorted current paths in Ag$_{2+\delta}$Te and Ag$_{2+\delta}$Se \cite{Parish2003}, which are narrow-gap semiconductors and are potentially TIs \cite{WZhang2012}; 
such distorted current paths could lead to negative MR, but only for the longitudinal configuration \cite{Hu2007}.  
Obviously, disorder in narrow-gap systems plays a key role in both positive and negative gigantic MR, and its understanding may help to engineer practical devices for magnetic data storage and sensors. 

\section{Methods}

{\bf Sample preparations.} 
Single crystals of TlBi$_x$Sb$_{1-x}$Te$_2$ were grown from a melt of high-purity (99.9999\%) starting materials of elementary Tl, Bi, Sb and Te shots that were cleaned
to remove the oxide layers formed in air \cite{Wang2015}. They were mixed in the target composition and sealed in evacuated quartz tubes. After heating to 600$^\circ$C for 48 h, at 
which the tubes were intermittently shaken to ensure homogeneity of the melt, they were slowly cooled down to 400$^\circ$C in 100 h. Typically the growth resulted in multiple intertwined 
crystallites whose size increases from bottom of the grown boule to the top, and single crystals suitable for further analysis were prepared by cleavage at the top part. 
The actual chemical composition of the samples was confirmed to be consistent with the nominal one by inductively-coupled plasma atomic-emission spectroscopy (ICP-AES) as well 
as by energy-dispersive X-ray spectroscopy (EDX).
The crystal structure was confirmed by powder X-ray diffraction to be unchanged from both endmembers \cite{Hockings1961} over the whole range of $x$.

\vspace{2mm}

{\bf Measurements.}
Transport measurements were performed on thin needle-like samples with a typical dimension of 4$\times$1 mm$^2$ in the $ab$ plane and 0.4 mm in the $c$ axis. 
Using a bath cryostat equipped with a variable-temperature insert the resistivity $\rho_{xx}$ and the Hall resistivity $\rho_{yx}$ were measured simultaneously  in sweeping magnetic fields ($\pm$14 T) with a low-frequency lock-in technique.
Electrical contacts were prepared in a standard six-probe configuration using 25 $\mu$m platinum wires attached with dissolved silver paint that was cured at room temperature. 

\vspace{2mm}

{\bf Numerical simulations.}
To simulate the behavior of the puddles in a magnetic field we have used the common static model of shallow donors and acceptors in compensated semiconductors \cite{Skinner2012,Skinner2013,Borgwardt2015}.
The donors with density $N_D$ and the acceptors with denity $N_D=K N_A$ are placed at random, uncorrelated positions. Their bare energies are seperated 
by a band gap $\Delta$ which was chosen as $15\times E_c$ in the simulations.

For a specific realization of the dopant positions (say $\mathbf{r}_1, \mathbf{r}_2, \dots$), the Hamiltonian reads
\begin{equation}\label{eq:ham_units}
 H = \frac{\Delta}{2} \, \sum_i f_i n_i +\frac{1}{2} \sum_{i \neq j} V_{ij} \; q_i q_j.
\end{equation}
$q_i$ denotes the charge in units of the elementary charge of the dopant at position $\mathbf{r}_i$. It can be either $0$ or $-1$ for acceptors ($f_i=-1$) and either $0$ or $+1$ for donors ($f_i=+1$).
The occupation $n_i$ of the $i$-th dopant is related to its charge $q_i$ by $q_i = (f_i+1)/2-n_i$.

The Coulomb interaction $V_{ij}$ between the dopants is equipped with a short distance cutoff of the order of the effective Bohr radius $a_B = 4\pi \varepsilon \varepsilon_0 \hbar^2/(m^* e^2)$,
where $m^*$ denotes the effective mass. This accounts for the finite extent of the dopant wavefunctions \cite{Skinner2012, Skinner2013}. With this approximation the effective interaction reads
$V_{ij}=\tfrac{e^2}{4 \pi \varepsilon \varepsilon_0 \sqrt{|\mathbf{r}_i-\mathbf{r}_j|^2+a_B^2}}$.
The model is valid for $\Delta \gg E_c= e^2/(4 \pi \varepsilon \varepsilon_0 N^{-1/3})$ (otherwise one has to include the band states) and $\varepsilon \gg 1$ to ensure that the dopants are shallow.

We have performed simulations for zero temperature. Only small temperature effects are expected as long as $T \ll E_c$ \cite{Borgwardt2015}.
To find the true groundstate is an exponentially hard problem, but there is an algorithm to find an approximate groundstate, called a pseudo-groundstate, in polynomial time \cite{ESbook, Skinner2012}.
The physical properties of a pseudo-groundstate are expected to be indistinguishable from that of the true groundstate.

We introduce the single electron energies as
$ \epsilon_j = \frac{\Delta}{2} f_j - \sum_{i \neq j} V_{i j} \; q_i$.
A pseudo-groundstate is found when $\delta E_{(i,j)} = \epsilon_j - \epsilon_i - V_{i j}>0$
is satisfied for all proper pairs with $n_j =0$ and $n_i = 1$. Simulations are performed in a cubic volume $V=L^3$ with periodic boundary conditions.
Once we have found a pseudo-groundstate, we plot the surface contour of constant potential as explained in the main text, where a typical result is shown.

For the calculation of the conductivity we proceed as follows once a pseudo-groundstate is found: 
The simulation volume is discretized and the grid points serve as nodes for the resistor network.
We calculate the local conductivity between two neighboring grid points from the Coulomb potential and the Zeeman energy as explained in the main text. 
The emerging resistor network is solved exactly, see Ref. \cite{Derrida1984} for details on the algorithm 
(however, in contrast to Ref. \cite{Derrida1984}, we use physical boundary conditions where the currents, rather than the potentials, are set to zero on the growing face of the network).
In this way, the conductivity of the full system is found. The grid size was chosen as $N^{-1/3}$, which is the typical lengthscale ($\sim$1.5 nm for $N = 3 \times 10^{20}$ cm$^{-3}$) on which the potential changes.
By using finer grids, we have checked that discretization effects are small ($\lesssim 5\%$ and well within the numerical error bars).
The simulations have been performed for $\approx 4.1 \times 10^5 $ dopants with periodic boundary conditions.
The orbital MR inside the puddles is neglected in the present calculations.


\vspace{5mm}

{\bf Acknowledgments} We thank D. I. Khomskii for helpful discussions.
This work was supported by DFG (CRC1238 ``Control and Dynamics of Quantum Materials", Projects A04 and C02).
O.B. acknowledges the support from Quantum Matter and Materials Program at the University of Cologne funded by the German Excellence Initiative.
The numerical simulations were performed on the CHEOPS cluster at RRZK Cologne.

{\bf Author contributions}
Y.A. conceived the project. Z.W., O.B., and A.A.T. performed the experiments.  J.L. performed numerical simulations which were interpreted by J.L. and A.R. O.B., J.L. and Y.A. wrote the manuscript with inputs from all authors.

{\bf Additional information}
Correspondence and requests for materials should be addressed to Y.A. (ando@ph2.uni-koeln.de)

\clearpage
\onecolumngrid

\renewcommand{\thefigure}{S\arabic{figure}} 
\renewcommand{\thesection}{S\arabic{section}.} 
\setcounter{figure}{0}
\setcounter{equation}{0}

\begin{flushleft} 
{\Large {\bf Supplementary Information for 
 ``Gigantic negative magnetoresistance in a disordered topological insulator''}}
\end{flushleft} 
\vspace{2mm}

\begin{flushleft} 
{\bf S1. Negative MR expected from 3D weak localization}
\end{flushleft} 

The TlBi$_x$Sb$_{1-x}$Te$_2$ (TBST) system studied here has a three-dimensional (3D) Fermi surface \cite{Trang2016}, and hence its transport should be treated as 3D. The magnetoconductance due to the weak localization effect in 3D systems has been calculated by Al'tshuler {\it et al.} \cite{Altshuler} as 
\begin{equation}
\Delta \sigma = \frac{e^2}{2 \pi^2 \hbar}\,  f_3\left(\frac{B}{B_{in}}\right)
\left(\frac{eB}{\hbar}\right)^{1/2},
\end{equation}
where $B_{in}$ is the characteristic magnetic field scale defined by the inelastic diffusion length $L_{\phi}$ via $B_{in} = \hbar / (4\pi e L_{\phi}^2)$, and $f_3(x) = 0.605$ for $x \gg 1$. For a typical $L_{\phi}$ of 1 $\mu$m,  $B_{in}$ = 0.329 mT and one can safely replace  $f_3(B/B_{in})$ with 0.605 in the relevant magnetic-field range. 
We have calculated the negative magnetoresistance (MR) expected from the 3D weak localization by using the above formula, and the result for 2 T is shown as an example in the main text.

\begin{figure*}[b]
\begin{center}
\includegraphics[width=16cm]{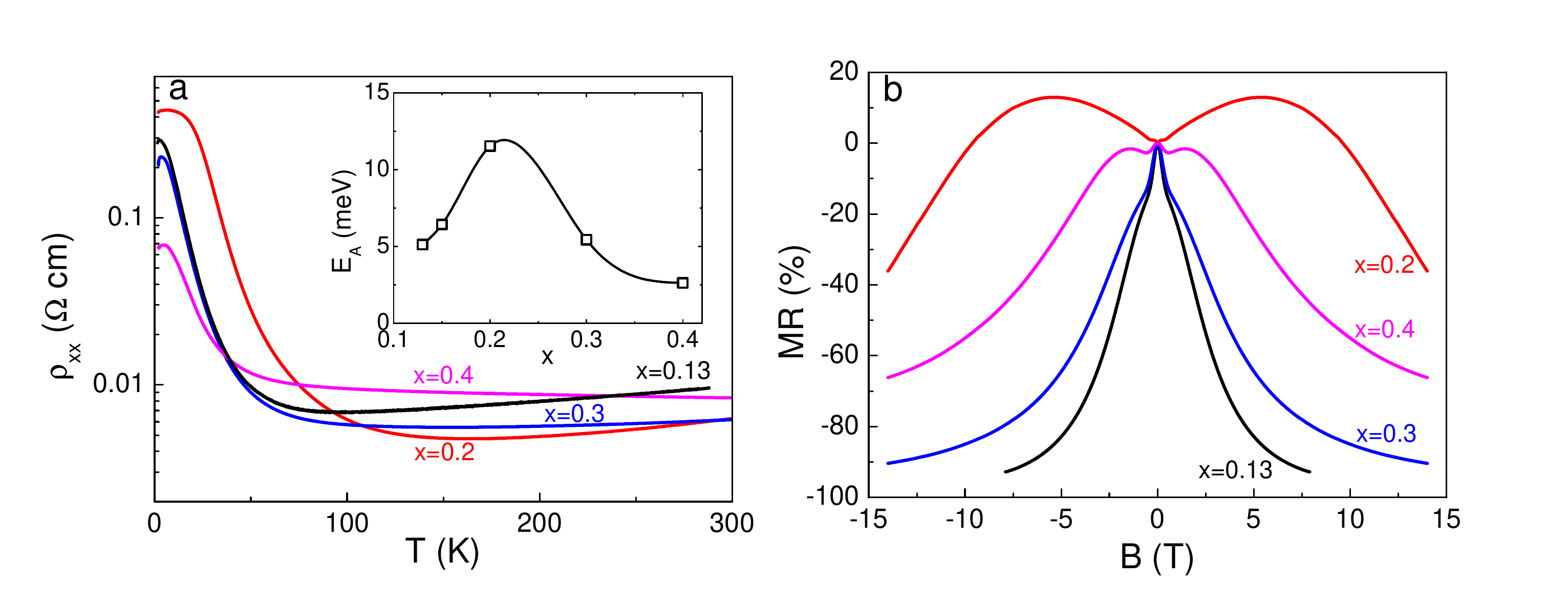}
\caption{
({\bf a}) Temperature dependencies of $\rho_{xx}$ in the $ab$ plane for $x$ = 0.13, 0.2, 0.3, and 0.4 in 0 T; inset shows the $x$ dependence of the effective activation energy extracted from the $\rho_{xx}(T)$ data (solid line is just a guide to the eye).
({\bf b}) Magnetoresistance of the same set of samples in the transverse configuration ($j \perp B \parallel c$) at 2 K.
} 
\end{center}
\end{figure*}

\begin{flushleft} 
{\bf S2. Resistivity and MR behavior at various $x$ values}
\end{flushleft} 

As is mentioned in the main text, all the samples of TBST that showed an insulating behavior ($0.13 \le x \le 0.4$) presented a negative MR, although the magnitude of the negative MR was smaller at $x$ values other than 0.15. Figure S1 shows the $\rho_{xx}(T)$ data and the MR data for $x$ = 0.13, 0.2, 0.3, and 0.4. The inset of Fig. S1a shows the effective activation energy extracted from the upturn in $\rho_{xx}(T)$ at low temperature; this effective activation energy is an indicator of the degree of compensation and presents a maximum at $x$ = 0.2. Interestingly, the negative MR at $x$ = 0.2 is smaller than that at $x$ = 0.15, which supports our interpretation that the gigantic negative MR requires the degree of compensation (quantified by the parameter $K$ in the main text) to be at a right value away from 1.0.

\newpage

\begin{flushleft} 
{\bf S3. Comparison of transverse and longitudinal MR}
\end{flushleft} 

When the MR is caused by the Zeeman effect as is proposed in the main text, it should be essentially isotropic as long as the $g$ factor is isotropic. We have performed experiments to make a direct comparison between transverse and longitudinal configurations using a $x$ = 0.15 sample which is different from the one shown in the main text. As one can see in Fig. S2, the anisotropy in the MR behavior measured at 2 K in three different orientations of the magnetic field are all similar, supporting our interpretation that the MR behavior stems from the Zeeman effect.

\begin{figure}
\begin{center}
\includegraphics[width=9cm]{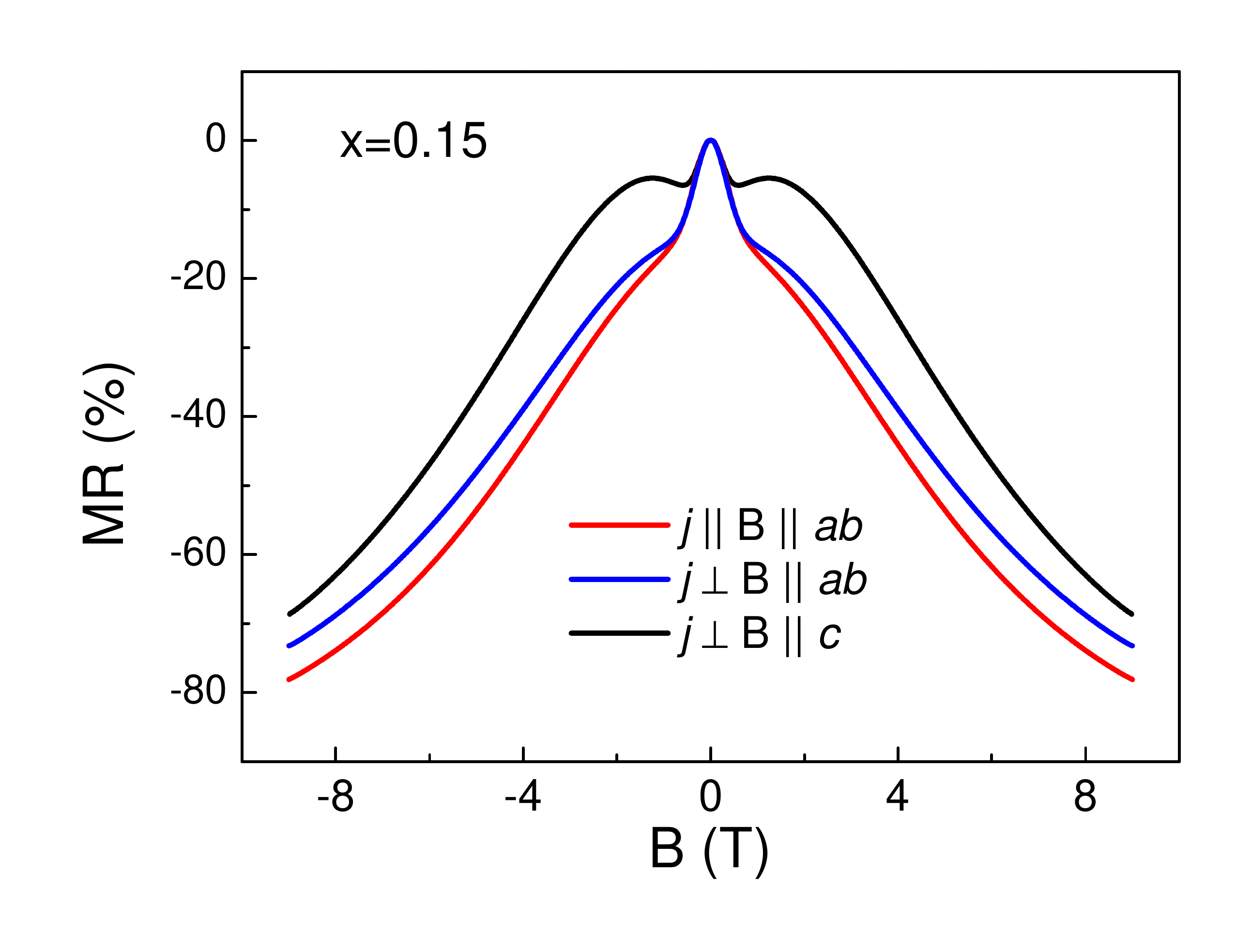}
\caption{
Magnetoresistance of another TlBi$_{0.15}$Sb$_{0.85}$Te$_2$ sample at 2 K measured in three different configurations,
longitudinal ($j \parallel B \parallel ab$), transverse in-plane ($j \perp B \parallel ab$), and transverse out-of-plane ($j \perp B \parallel c$). 
} 
\end{center}
\end{figure}


\end{document}